\documentclass[aoas]{imsart}

\RequirePackage{amsthm,amsmath,amsfonts,amssymb,tikz,bbm}
\RequirePackage[authoryear]{natbib}
\RequirePackage[colorlinks,citecolor=blue,urlcolor=blue]{hyperref}
\RequirePackage{graphicx}
\graphicspath{ {./Images/} }
\usetikzlibrary{positioning} 
\startlocaldefs
\theoremstyle{plain}

\theoremstyle{remark}

\endlocaldefs

\begin{document}

\begin{frontmatter}
\title{Averaging polyhazard models using Piecewise deterministic Monte Carlo with applications to data with long-term survivors}
\runtitle{Averaging polyhazard models using Piecewise Deterministic Monte Carlo}

\begin{aug}
%%%%%%%%%%%%%%%%%%%%%%%%%%%%%%%%%%%%%%%%%%%%%%%
%% Only one address is permitted per author. %%
%% Only division, organization and e-mail is %%
%% included in the address.                  %%
%% Additional information can be included in %%
\author[A]{\fnms{Luke}~\snm{Hardcastle}\ead[label=e1]{luke.hardcastle.20@ucl.ac.uk}},
\author[A]{\fnms{Samuel}~\snm{Livingstone}\ead[label=e2]{samuel.livingstone@ucl.ac.uk}}
\and
\author[A]{\fnms{Gianluca}~\snm{Baio}\ead[label=e3]{g.baio@ucl.ac.uk}}
%%%%%%%%%%%%%%%%%%%%%%%%%%%%%%%%%%%%%%%%%%%%%%
%% Addresses                                %%
%%%%%%%%%%%%%%%%%%%%%%%%%%%%%%%%%%%%%%%%%%%%%%
\address[A]{Department of Statistical Science,
University College London\printead[presep={,\ }]{e1,e2,e3}}

\end{aug}

\begin{abstract}
Polyhazard models are a class of flexible parametric models for modelling survival over extended time horizons. Their additive hazard structure allows for flexible, non-proportional hazards whose characteristics can change over time while retaining a parametric form, which allows for survival to be extrapolated beyond the observation period of a study. Significant user input is required, however, in selecting the number of latent hazards to model, their distributions and the choice of which variables to associate with each hazard. The resulting set of models is too large to explore manually, limiting their practical usefulness. Motivated by applications to stroke survivor and kidney transplant patient survival times we extend the standard polyhazard model through a prior structure allowing for joint inference of parameters and structural quantities, and develop a sampling scheme that utilises state-of-the-art Piecewise Deterministic Markov Processes to sample from the resulting transdimensional posterior with minimal user tuning.  
\end{abstract}


\end{frontmatter}

\section{Introduction}
\label{sec;Introduction}

Polyhazard models are a class of flexible parametric models for time-to-event data, defined by additively combining hazards from simpler, typically one- or two-parameter survival distributions
\begin{equation*}
    h_Y(y) = \sum_{j=1}^Kh_j(y),
\end{equation*}
where $Y$ is a random variable representing a time-to-event outcome, and $h_Y(y), h_j(y)$ are hazard functions. 

This additive procedure results in hazard functions that are flexible and able to model a wide-range of covariate effects. Originally developed for analysis of latent competing risks \citep{Berger1993, Louzada1999}, polyhazard models have become increasingly popular for modelling long-term survival required for Health Technology Assessment (HTA) following the work of \cite{Demiris2015}, who used a poly-Weibull model to analyse survival in transplant patients, and the inclusion of polyhazard models in an influential UK National Institute for Health and Care Excellence (NICE) review \citep{Rutherford2020}.

HTA provides a decision-theoretic framework for healthcare systems to assess the cost-effectiveness of medical interventions, with new interventions being introduced when their expected benefit sufficiently outweighs their expected cost in comparison to the current standard of care. When these benefits are realised over a patient's lifetime this requires the computation of
\begin{equation}
\label{eq;MeanSurv}
    \mathbb{E}[Y] = \int_0^{\infty}S_Y(y)dy, \quad S_Y(y) = \mathbb{P}(Y > y).
\end{equation}

This is in contrast to common approaches to analysing clinical trials, which typically target measures such as \textit{median} survival, and confounds standard non-parametric tools (e.g Kaplan-Meier estimates) when patients are still alive at the end of the observation period as they contain no mechanism to extrapolate beyond final event times. This is invariably the case in medical applications due to patient drop-out and the financial and ethical constraints that limit the length of both clinical trials and observational studies. Alternative approaches are therefore required to infer long-term survival.

A common approach is to impose parametric assumptions on $Y$, which given parameters $\boldsymbol\theta$, allows \eqref{eq;MeanSurv} to be computed either analytically or through a simple numerical approximation, with or without censored observations. This broadly follows the recommendations of \cite{Latimer2011} who proposes a set of two- or three-parameter survival distributions to be used for this purpose; given the leading role, globally, of NICE, these have become the gold-standard in HTA. While parsimonious, these distributions are typically restricted to hazards that are increasing, decreasing or unimodal and covariate effects restricted by assumptions of proportional hazards or odds. Further, these standard models infer the parameters dictating extrapolation from the whole sample, while in reality observations at the end of the trial are likely to contain more information about how survival can be expected to evolve in the long-term.

Polyhazard models can capture a much wider range of hazard curves while retaining the interpretability and parsimony of simpler model. Further, due to the additive decomposition of the hazard function, later observations naturally have more influence on long-term survival. 

Despite theses advantages applications of polyhazard models have been limited due to: \textit{i)} the lack of accessible computational tools and understanding of how prior specification affects inference; \textit{ii)} a number of structural choices which, in the presence of even a small number of covariates, leads to a space of candidate models which is infeasibly large to explore manually. 

\subsection{Motivating datasets}

This paper is motivated by the application of polyhazard models to compute mean survival in two datasets where the presence of long-term survivors necessitates the need for extrapolation. In both settings application of polyhazard models would have previously been infeasible due to the issues outlined at the end of the preceding Section.

The first, the Copenhagen Stroke Study \citep[COST;][]{Jørgensen1996}, contains survival times for stroke survivors with 13 relevant covariates. Previous works have used this study to investigate the long-term risks faced by stroke survivors. \cite{Kammersgaard2004} sought to understand the prognosis for very old patients (defined as  $\text{age} \geq 85$), conducting a subanalysis using Cox proportional hazards regression with very old age, stroke severity score and presence of atrial fibrilation as covariates. \cite{Andersen2005} investigated the association between sex and survival outcomes, fitting a Cox proportional hazards model to artificial 1-, 5- and 10- year data cuts to assess the changing effect of sex on survival in the short- and long-term. Similarly, \cite{Andersen2011} investigated the interaction between stroke severity, as defined by the stroke severity score, and other prognostic indicators. 

The second contains survival times following Kidney transplantation in Taiwan \citep{Chen2022a}. The original analysis used hazard ratios provided by a Cox regression to understand the impact of transplant waiting times on long-term survival. Patients were split into four groups based on wait times ($<$1 year, 1-3 years, 3-6 years, $>$6 years). Additional covariates in the data include age at time of transplantation (defined in 10 year blocks), sex, hypertension and Dyslipidemia. The primary challenge with the analysis of these data are the high censoring rates in all age and waiting time groups, with only the oldest patient group (71-80 years) reaching median survival with 41.18\% censored, and censoring rates of 89.90\% and 92.00\% in the youngest two age groups.

\subsection{Paper structure}

This paper is concerned with addressing these issues via Bayesian model averaging, to allow the application of polyhazard models to the motivating data. In Section \ref{sec;Polyhazard} we extend the polyhazard model by accounting for uncertainty in structural choices through an extended prior specification leading to a Bayesian model averaging approach. In Section \ref{sec;MCMC} we develop bespoke Markov Chain Monte Carlo (MCMC) methodology extending existing sampling methods based on piecewise deterministic Markov processes \citep[PDMPs;][]{Fearnhead2018}.  This allows for efficient generation of posterior samples, reducing the computational burden from fitting each individual polyhazard model to fitting a small set of models with high posterior mass. PDMP-based samplers have emerged as a promising new direction in Bayesian computation. Their development has been hindered, however, by a limited understanding of their effectiveness in applied settings, a limitation this paper begins to address. In Section \ref{sec;Applications} we study the extended model by re-analysing a digitised version of data first studied by \cite{Demiris2015}. Through this comparative analysis we show the effect non-informative vs weakly informative priors in this setting and the importance of accounting for structural uncertainty. Following this we apply the extended polyhazard model to the motivating data.

\section{Polyhazard models} 
\label{sec;Polyhazard}
\subsection{Survival analysis}
\label{sec;Survival}
Survival analysis involves the study of the time until an event of interest (e.g death or cancer progression), denoted $Y$. In this paper we will assume we have observed independent samples $\mathcal{D} = (y_i, c_i, \boldsymbol{x}_i)_{i=1}^n$. Here $y_i \in \mathbb{R}_+$ are such that if $c_i = 1$, then $Y_i = y_i$, i.e $y_i$ is the observed time the event occurred. Conversely if $c_i = 0$, then $Y_i > y_i$, and we refer to individual $i$ as having been (right-)censored. Commonly this is due to patient dropout or the patient surviving beyond the end of the study period. We assume non-informative censoring, whereby the censoring mechanism is independent of the event of interest. Finally, $\boldsymbol{x}_i \in \mathbb{R}^p$ is a $p-$dimensional vector of covariates for individual $i$.

Survival analysis is typically concerned with the study of the hazard function $h(y)$, representing the instantaneous risk of the event occurring, and the survivor function $S(y)$, where
\begin{equation}
\label{eq;functions}
    h(y) := \lim_{\varepsilon\to0}\frac{\mathbb{P}(Y \leq y + \epsilon \mid Y > y)}{\varepsilon}, \quad S(y) := \mathbb{P}(Y > y) = \exp\left(-\int_0^yh(u)du\right).
\end{equation}
In the parametric setting, given parameters $\boldsymbol\theta$, these quantities are combined to form a likelihood
\begin{equation*}
    \mathcal{L}(\boldsymbol\theta\mid\mathcal{D}) = \prod_{i=1}^nh_{\boldsymbol\theta}(y_i)^{c_i}S_{\boldsymbol\theta}(y_i).
\end{equation*}
Note that given the correspondence between survivor and hazard functions in \eqref{eq;functions} this implies that selecting the hazard function corresponds directly to model selection.

\subsection{Polyhazard model definition}
\label{sec;PolyhazardDef}
Polyhazard models \citep{Berger1993, Louzada1999} are constructed by combining multiple independent parametric hazards via the additive formulation
\begin{equation}
\label{eq;add_haz}
    h_{\boldsymbol{D},\boldsymbol\theta, \boldsymbol\gamma}(y\mid \boldsymbol{x})  = \sum_{k=1}^Kh_{D_k,\boldsymbol\gamma_k,\boldsymbol\theta_k}(y\mid \boldsymbol{x}).
\end{equation}

Each subhazard corresponds to a proper hazard function from a known distribution $D_k$, selected from a set of candidate distributions $\mathcal{H}$, and with subhazard specific parameters $\boldsymbol\theta_k$. Each $\boldsymbol\theta_k$ is composed of a shape parameter $\nu_k$, and rate,\footnote{We will avoid the term rate parameter, to prevent confusion with the rates required for posterior sampling, outlined in Section \ref{sec;MCMC}.} scale or location parameter $\mu_k$, such that $\boldsymbol\theta_k = (\nu_k, \mu_k)$. For each hazard, covariate information is included in the location parameter via a log-link, such that
\begin{equation}
\label{eq;lin_pred}
    \mu_k(\boldsymbol{x},\boldsymbol\gamma_k) = \exp\left(\beta_{k,0} + \sum_{j:\gamma_{kj} =1}x_{j}\beta_{k,j}\right),
\end{equation}
where $\gamma_{kj} \in \{0,1\}$ indicates whether the $j^{th}$ covariate is included in the $k^{th}$ subhazard. In practice we will centre and normalise each element of $\boldsymbol{x}$ such that for a given hazard $\beta_{k,0}$ can be interpreted as the location parameter for the average individual in the sample. This information is collated as $\boldsymbol{D} = (D_k)_{k=1}^K$, $\boldsymbol\theta = (\boldsymbol\theta_k)_{k=1}^K$, and $\boldsymbol\gamma = (\boldsymbol\gamma_k)_{k=1}^K$, such that the model is completely defined by the specification of $(K, \boldsymbol{D}, \boldsymbol\gamma, \boldsymbol\theta)$. 

We place no restriction on the combination of simpler hazard forms, neither requiring each subhazard to be from the same parametric family nor requiring each parametric family to be represented in \eqref{eq;add_haz}. Similarly, $\boldsymbol\gamma_k$, need not be identical across all subhazards.

In this paper we will focus on polyhazard models where $\mathcal{H}$ contains the Weibull and log-logistic distributions with respective hazard functions
\begin{equation*}
    h_W(y) = \mu\nu y^{\nu-1}, \quad h_{LL}(y) = \frac{(\frac{\nu}{\mu})(\frac{y}{\mu})^{\nu-1}}{1 + (\frac{y}{\mu})^\nu},
\end{equation*}
while noting that the methods presented naturally extend to other choices \citep[see for example][]{Louzada1999}. 

\begin{figure}
    \centering
    \includegraphics[width=5.5in]{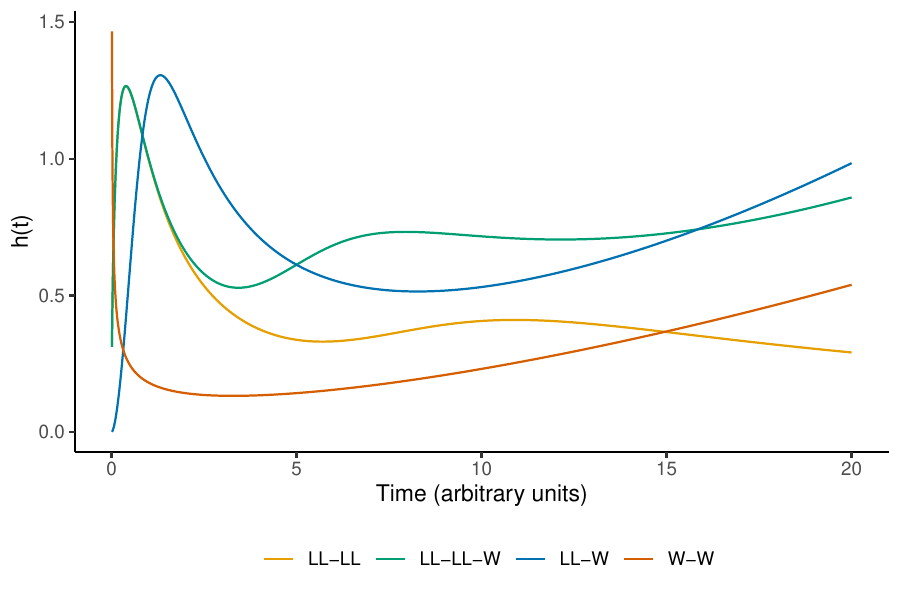}
    \caption{Example hazard shapes obtainable by the polyhazard model with combinations of log-logistic (LL) and Weibull (W) latent hazards.}
    \label{fig:2Hazards}
\end{figure}

Combining hazard functions with different shapes results in flexible baseline hazards and covariate effects that are more flexible than those possible with simpler models. Various example hazard shapes generated by combining  Weibull and log-logistic hazards are shown in Figure \ref{fig:2Hazards}. 

We briefly address two common misconceptions regarding the polyhazard model. \textit{i)} The polyhazard model is \textit{not} a mixture model. In contrast, each individual in the population is subject to risk from every subhazard (with intensity determined by relevant covariates), and there is no explicit weighting of subhazards within the population. \textit{ii)} While the form of \eqref{eq;add_haz} is recognisable as the hazard for an individual subjected to independent, latent competing risks, we do not necessarily assume that the data were generated in this way. Rather, we utilise the form of \eqref{eq;add_haz} as a flexible modelling assumption.  

Standard application of polyhazard models typically follows one of two approaches. In the first $K,\boldsymbol{D}$ and $\boldsymbol\gamma$ are fixed \textit{a priori}, meaning inference is performed on $\theta$ only. In HTA applications, for example, it has become common to only consider the bi-Weibull model (e.g \citealt{Negrin2017}). This often means that potentially viable candidate models are excluded from the analysis without justification. Alternatively, $K,\boldsymbol{D}$ and $\boldsymbol\gamma$ are reduced to a small set of possible values for which all models are fitted and compared \textit{a posteriori}.  \cite{Demiris2015} compare poly-Weibull models with $K$ from 1 to 4 and $\gamma$ based on the deviance and clinical plausibility, and \cite{Benaglia2015} compare the bi-Weibull and bi-Gompertz model based on visual fit.

Both these approaches rely on the set of candidate models being small enough to fit and interrogate individually, which is very restrictive, as for a fixed maximum number of subhazards, $K_{\max}$, the size of the set of candidate models is given by
\begin{equation*}
    \sum_{k=1}^{K_{\max}}2^{pk}{{|\mathcal{H}| + k -1}\choose{k}}.
\end{equation*}
The result is a model space which is infeasible to explore manually for anything beyond very small $K_{\max}$, $\mathcal{H}$ and $p$. For the kidney transplant data analysed in Section \ref{sec;TKT}, taking $K_{\max} = 3$, $|\mathcal{H}| = 2$ and $p=13$, results in 274,928,246,784 candidate models.

\subsection{Priors}
\label{sec;Priors}
We now introduce an extended specification of prior for the polyhazard model, which will incorporate uncertainty across each element of $(K, \boldsymbol{D}, \boldsymbol\gamma, \boldsymbol\theta)$. This induces posterior model weights that can then be used for Bayesian model selection or averaging. The prior, denoted throughout by $\pi_0(\cdot)$, is specified as
\begin{equation*}
    \pi_0(K, \boldsymbol{D}, \boldsymbol\gamma, \boldsymbol\theta, \boldsymbol\phi) \propto \pi_0(\boldsymbol\theta \mid K,\boldsymbol{D},\boldsymbol\gamma, \boldsymbol\phi)\pi_0(\boldsymbol\gamma \mid K, \boldsymbol\phi)\pi_0(\boldsymbol\phi)\pi_0(\boldsymbol{D})\pi_0(K),
\end{equation*}
where $\boldsymbol\phi = (\omega, \sigma_\beta)$ is a vector of hyperparameters to be defined. 

First considering $\boldsymbol\theta \mid K, \boldsymbol{D}, \boldsymbol\gamma, \boldsymbol\phi$, we specify
\begin{gather*}
    \log(\nu_k) = \alpha_k \sim \text{Normal}(0,\sigma_\alpha), \quad k = 1,\dots,K, \\
    \beta_{k,0} \sim \text{Normal}(0,\sigma_{\beta_0}), \quad k = 1,\dots,K.
\end{gather*}

We place weakly informative priors on each $(\nu_k, \beta_{k,0})$ independent of distribution, setting $\sigma_\alpha = 2$ following the reasoning in \cite{Demiris2015}, and $\sigma_{\beta_0} = 5$. We provide further justification for, and discussion of, this choice in Section \ref{sec;Prior_sens}, but briefly note that this specification must account for the scale of the data (years in all the applications presented in Section \ref{sec;Applications}). A contrasting approach is taken for the poly-Weibull model by \cite{Demiris2015} and \cite{Benaglia2015} who place a $\text{Uniform}(0,1)$ prior on $\nu_1$. While justifiable for fixed $(K, \boldsymbol{D})$, the effect of this prior on the posterior is unclear when $(K, \boldsymbol{D})$ are also being inferred.

For the remaining linear predictor terms in \eqref{eq;lin_pred} we account for uncertainty in the effect of the covariates on the outcome through the specification of the spike-and-slab prior \citep{Mitchell1988}
\begin{gather}
\label{eq;SnS}
    \pi_0(d\beta_{k,j}\mid\boldsymbol\phi) \propto (1-\omega)\delta_0(d\beta_{k,j}) + \omega \Tilde{\pi}_0(\beta_{k,j}\mid \sigma_\beta), \\
    \nonumber
    k = 1,\dots,K, \quad j = 1,\dots,p,
\end{gather}
where $\Tilde{\pi}_0(\cdot\mid\sigma_\beta)$ is the density of a Normal distribution with mean 0 and $\delta_0$ is a Dirac measure centred at 0.  This formulation implies independent $\text{Bernoulli}(\omega)$ priors for each element of $\gamma_k$, resulting in 
\begin{equation*}
    \pi_0(\boldsymbol\gamma\mid K) \propto \omega^{\sum_{k,j}\gamma_{k,j}}(1-\omega)^{pK - \sum_{k,j}\gamma_{k,j}},
\end{equation*}
and we extend this to a hierarchical modelling setting through a conjugate Beta prior on $\omega$,
\begin{equation*}
    \omega \sim \text{Beta}(a,b),
\end{equation*}
as recommended by \cite{Kohn2001}. This is a well established approach, which reduces the influence of prior specification in the context of Bayesian model averaging \citep{Ley2009}. When applied to the COST and kidney transplant data we set $a = b = 4$. Further to this, in order to regularise the effect sizes observed in the linear predictors we utilise a horseshoe, half-Cauchy hyperprior on $\sigma_\beta$,
\begin{equation*}
    \sigma_\beta \sim \text{Cauchy}_{>0}(0,1),
\end{equation*}
designed to circumvent well known model misspecifcation issues arising from using a fixed $\sigma_\beta$ \citep{Polson2012}.

Note that in the above formulation $\omega$ and  $\sigma_\beta$ are shared hyperparameters across subhazards encouraging sharing of information between hazards about expected effect sizes, which implies that the induced prior on $|\boldsymbol\gamma|$ should be interpreted as a prior on the number of covariates across the model, rather than the number of covariates associated with each individual subhazard. 

Each subhazard distribution, $D_k$, is drawn uniformly from the set of candidate distributions
\begin{equation*}
    D_k \mid K \sim \text{Uniform}(\mathcal{H}),
\end{equation*}
inducing a multinomial prior on $D$. If expert knowledge favours certain subhazards being present in the model this can be encoded at this stage.

Finally, prior belief about the number of hazards in the model is represented through a truncated Poisson prior
\begin{equation*}
    K \sim \text{Poisson}_{>0}(\xi),
\end{equation*}
for fixed $\xi$. We set $\xi = 2$ defining a weakly informative prior, encoding a soft preference for models with a smaller number of hazards. Any discrete distribution could be used as, for example, there may be expert knowledge which suggests a strong prior belief that $K > 2$, however in practice we find there is rarely justification for $K>4$. This is reflected in the choice of $\xi$ which implies $\mathbb{P}(K > 4) = 0.061$ \textit{a priori}.

\section{Posterior sampling}
\label{sec;MCMC}

The posterior induced by the prior formulation of Section \ref{sec;Polyhazard} presents a challenging target distribution for many of the standard posterior sampling tools of Bayesian inference. Difficulties stem from the varying dimension of the parameter space and changing form of the likelihood due to the priors on $(K, \boldsymbol{D}, \boldsymbol\gamma)$, as well as the geometry of the posterior when $(K,\boldsymbol{D},\boldsymbol\gamma)$ are fixed. Here, when the data are highly censored, the marginal posteriors of parameters for subhazards (which are influential later in the follow-up period) are often skewed due to partial information from censored observations. Further, subhazards can switch roles in the model. When these subhazards are from the same distribution, exchangeable prior information results in a symmetric, multimodal  posterior with $K!$ modes. Role switching, however, can also occur when the subhazards have different distributions, inducing a non-symmetric, multimodal posterior. This is akin to the label switching problem in mixture models (e.g. \cite{Jasra2005}).  An example is shown in the supplementary materials and we discuss this issue further in Section \ref{sec;Output}. In this Section we develop a bespoke sampling algorithm to handle these challenging posterior features.

Current approaches to posterior computation for fixed $(K, \boldsymbol{D}, \boldsymbol\gamma)$ include a Gibbs sampler implemented in WinBUGS and a Stan implementation of the No-U-Turn Sampler, both for the poly-Weibull model \citep{Demiris2015, Baio2020}. Neither of these approaches naturally extend to the transdimensional case. The former is also susceptible to high levels of auto-correlation, while both can struggle in the presence of multimodality.

The foundation of the method developed in this Section is the Zig-Zag sampler \citep{Bierkens2019}, an example of a class of novel MCMC methods based on continuous-time piecewise deterministic Markov processes \citep[PDMPs; ][]{Fearnhead2018}. These processes are non-reversible, helping navigate some of the challenging geometry of the posterior \citep{Andrieu2021}, and are able to use their continuous, piecewise deterministic sample paths to directly sample from spike and slab distributions \citep{Chevallier2023, Bierkens2023} as defined by \eqref{eq;SnS}. These continuous time dynamics are combined with jump processes for updating $(K, \boldsymbol{D}, \boldsymbol\phi)$, allowing navigation of the full posterior.

\subsection{Zig-Zag sampling}
\label{sec;ZZS}
\begin{figure}
    \centering
    \includegraphics[width = 5.5in]{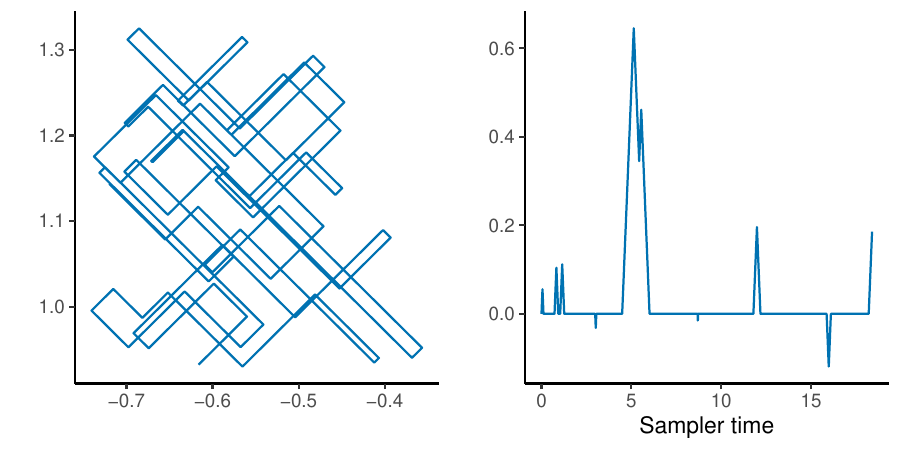}
    \caption{Trajectories from the Zig-Zag sampler (left) and variable selection Zig-Zag sampler (right).}
    \label{fig:ZZS}
\end{figure}

We focus for the moment on sampling the parameters $\boldsymbol\theta \in \mathbb{R}^{2K + |\boldsymbol\gamma|}$ conditional on fixed $(K, \boldsymbol{D}, \boldsymbol\gamma, \boldsymbol\phi)$. The Zig-Zag sampler augments $\theta$ with unit velocities, $\boldsymbol{v} \in \{-1,1\}^{2K + |\boldsymbol\gamma|}$, which define the deterministic, continuous-time evolution of $\theta$, such that 
\begin{equation*}
    \boldsymbol\theta_{s+t} = \boldsymbol\theta_s + v_st.
\end{equation*}
This process is interrupted by coordinate-wise velocity flips which update the $i^{th}$ component of $\boldsymbol{v}$ at time $t$ by $v_{t,i} \mapsto -v_{t,i}$. Writing the posterior as $\pi(\boldsymbol\theta) \propto \exp(-U(\boldsymbol\theta))$, where $U$ is commonly referred to as the potential \citep{Faulkner2024}, $v_{t,i}$ is flipped at times given by an inhomogeneous Poisson process (IHPP) with rate 
\begin{equation}
\label{eq:ZZS}
    \Lambda^F_i(t) = \max\{0, v_{t,i}\partial_iU(\theta_{t,i})\}.
\end{equation}
Intuitively, in the $i^{th}$ coordinate, if the process is moving into areas of lower potential (equivalently higher posterior density) it continues uninterrupted. If, however, the converse is true, then $v_{t,i}$ flips with rate proportional to the rate of growth in the potential. The result is an almost-surely continuous (on $\theta$-space), piecewise deterministic process, whose sample paths produce a zig-zag pattern as shown in Figure \ref{fig:ZZS} (left).

In practice, to avoid simulating $d$ IHPPs, the next event time is generated from $\Lambda^F(t) = \sum_{i=1}^d\Lambda^F_i(t)$. Once an event time $\tau_F$ has been generated the coordinate to switch is then chosen with probability proportional to $\Lambda^F_i(\tau_F)$. See \cite{Fearnhead2018} for a more detailed discussion.

\subsubsection{Generating the inhomogeneous Poisson process}
\label{sec;IHPP}
The efficiency of the Zig-Zag sampler is crucially dependent on the cost of generating event times from an IHPP with rate $\Lambda^F(t)$. This is most commonly achieved via Poisson thinning \citep{Lewis1979}, in which a proposed event time $\tau_F$ is generated from a dominating Poisson process with rate $M(t) > \Lambda^F(t)$, accepted with probability $\Lambda^F(\tau_F)/M(\tau_F)$; if the proposed move is rejected, the process continues with the same dynamics from time $\tau_F$.

While it is possible to derive a tight upper bound analytically in some cases, we know of no such choice of $M(t)$ that is suitable for polyhazard models. We therefore numerically bound $\Lambda^F(t)$ on the interval $[t_0, t_0 + t_b)$, via an extension of the Automatic Zig-Zag method  of \cite{Corbella2022}.As $t_b \to 0$,  the resulting $M(t)$ becomes arbitrarily tight meaning that proposed events are accepted with probability close to 1 (meaning there is little to no thinning), but at the cost of having to compute $M(t)$ over a large number of intervals before an event is observed. The computational efficiency of the sampler is therefore dependent on balancing the cost of constructing a tighter upper bound $M(t)$ against that of rejecting too many proposed events when the bound is loose.%  of computing $M(t)$ against the number of rejected events.

In the Automatic Zig-Zag approach a constant upper bound for $\Lambda^F(t)$ is found using Brent's method on an interval with fixed length. Costly, repeated gradient evaluations are avoided by performing a monotonicity check after the first iteration, which if passed allows the evaluation of $\Lambda^F(t)$ at one end of the interval to be used as the bounding rate.

We extend this approach in the following ways. Firstly, we replace the first iteration of Brent's method with evaluations of $\Lambda^F(t)$ at $\{t_0, t_0 + t_{b}/2, t_0 + t_{b}\}$. We use these evaluations to check monotonicity \emph{and} convexity. If both these checks are passed we then use the linear bound 
\begin{equation*}
    M(t) = \Lambda^F(t_0) + \frac{\Lambda^F(t_0 + t_b) - \Lambda^F(t_0)}{t_b}t, \quad t \in [t_0, t_0 + t_b),
\end{equation*}
which is provably tighter than the constant choice. If monotonicity holds but convexity does not we use the relevant evaluation at the end of the interval as a constant upper bound, and if neither hold we resort to Brent's method. In both of the latter two instances the resulting bound is as in the Automatic Zig-Zag, but when it is applicable we have found that the linear bound can be much tighter than a constant choice, which can speed up the sampler significantly.

The second modification is to adaptively set the length of the bounding interval, as has previously been suggested in a similar context by \cite{Sutton2023}, who recommend setting the length of the interval $t_b$ to be the 80\textsuperscript{th} percentile of observed inter-event times, $t^*$. We extend this approach to set $t_b = \min\{t^*, \Lambda^F(t_0)^{-1}\}$, which uses information from both the history and current state of the chain. Intuitively, if the evaluation of the rate is high at the current state of the chain, a shorter interval is likely to be appropriate. This heuristic is regularised by $t^*$ to avoid long intervals induced by small $\Lambda^F(t_0)$, which are likely to result in inefficient bounds. We note that in contrast to many adaptive MCMC schemes, this approach does not change the law of the process, and therefore we do not need to make considerations such as diminishing adaptation \citep[e.g.][]{Andrieu2008}.

Finally, we introduce a constant offset rate $\Lambda_0$ which is added to $M(t)$ to offset numerical errors and any failures in the checks described above. If the bounding does fail, this is easily diagnosed by reporting instances when the upper bound is exceeded. These errors can then be investigated or the offset increased.

\subsection{Updating hyperparameters}
\label{sec;Hyperparameters}
The hyperparameters $(\omega,\sigma_\beta)$ could be sampled directly by the Zig-Zag sampler, but strong posterior dependence between parameters and hyperparameters induced by the hyperprior structure would inhibit sampling efficiency.  A more elegant solution is to follow the Gibbs Zig-Zag approach of \cite{Sachs2023}, which allows traditional Gibbs updates to be interwoven into the Zig-Zag sampler at exponentially distributed intervals with rate $\Lambda^H$. In particular this allows $\omega$ to be updated by the closed form full conditional due to the Beta-Binomial prior formulation. 

Full conditionals for $\sigma_\beta$ are not available in closed form. However, sampling can be performed via adaptive random walk Metropolis steps. To avoid sampling difficulties resulting from the heavy-tails of the Cauchy distribution we utilise the re-parameterisation proposed by \citet{Betancourt2018}
\begin{gather*}
    \sigma_\beta = z_1\sqrt{z_2}, \\
    z_1 \sim \text{Normal}(0,1), \quad z_2 \sim \text{Inv-Gamma}(1/2,1/2),
\end{gather*}
and determine the step-size and covariance matrix of the random walk Metropolis proposal adaptively using a Robbins-Monro style updating scheme as seen in Algorithm 4 of \cite{Andrieu2008}.

\subsection{Zig-Zag sampling for variable selection}
\label{sec;VS}
Bayesian variable selection is a challenging problem even in standard parametric survival models. Current state-of-the-art approaches involve focusing sampling efforts on the marginal posterior for the variable inclusion indicator $\pi(\boldsymbol\gamma)$, where $\boldsymbol\gamma \in \{0,1\}^p$. Efficient exploration of the state space, however, requires efficient approximations of the marginal likelihood, which are typically not straightforward for polyhazard models \citep{Liang2023}. Furthermore, simpler, uninformed schemes such as the add-delete-swap reversible jump sampler of \cite{Newcombe2017} are likely inhibited by poor acceptance rates.

An alternative approach, concurrently developed by \cite{Chevallier2023} and \cite{Bierkens2023}, is to utilise the continuous sample paths of the Zig-Zag sampler to directly sample from the spike and slab posterior induced by \eqref{eq;SnS}. Here the process sticks to the hyperplane $\{\beta_{k,j} = 0\}$, corresponding to the spike, whenever it crosses it, by setting the corresponding velocity to 0 and then resetting the velocity after a waiting time, $\tau_\beta$. Specifying $\tau_\beta$ as the first time of the homogeneous Poisson process
\begin{equation*}
    \Lambda^{V}_{k,j}(t) = \frac{\omega}{1-\omega},
\end{equation*}
preserves the correct target distribution. We note that the key point of this construction is that the rate of unsticking is given by the posterior ratio between the models with $\gamma_{kj} = 1$ and $\gamma_{kj} = 0$. Since this ratio is being evaluated at $\beta_{k,j} = 0$, the likelihood takes the same value for $\gamma_{kj} = 1$ and $\gamma_{kj} = 0$, and this ratio cancels to a ratio of priors resulting in a homogeneous Poisson process. This approach, therefore, has the dual advantage of being informed by the current state of the process and also being computationally efficient as updates to $\gamma$ do not require any likelihood or gradient evaluations beyond those required for sampling $\theta$. Example trajectories for this process are given in Figure \ref{fig:ZZS} (right).

We extend the work of \cite{Chevallier2023, Bierkens2023} by including a hyperprior structure on $\omega$ as detailed in Section \ref{sec;Priors}. Directly sampling $\omega$ via the Zig-Zag sampler would result in unsticking times given by an inhomogeneous Poisson process requiring additional computational cost to generate. Alternatively by updating $\omega$ with a continuous-time jump process as described in Section \ref{sec;Hyperparameters}, the waiting times remain easy to generate as the first time of a Poisson process with piecewise constant rate.

\subsection{Birth-death-swap processes}
\label{sec;BDS}
The final sampling ingredient is a birth-death-swap process which is able to update the number of hazards $K$ and the vector of subhazard distributions $D$ in continuous time. Births, deaths and swaps occur at rates given by $\Lambda^b(t)$, $\Lambda^d(t)$ and $\Lambda^s(t)$ respectively, with corresponding proposal distributions for new parameters given by $q_b(\boldsymbol{u})$, $q_d(\boldsymbol{u})$ and $q_s(\boldsymbol{u})$. We note that in addition to allowing exploration of the posterior for $(K,\boldsymbol{D})$, these transdimensional updates also allow for traversal between modes for fixed $(K,\boldsymbol{D})$.

\subsubsection{Birth-death process}
\label{sec;BD}
To define the birth-death process we require that a detailed balance condition is met
\begin{equation}
\label{eq;BD_balance}
    \Lambda^b(t)\pi(\boldsymbol\theta, \boldsymbol{D}, K)q_b(\boldsymbol{u}) = \Lambda^d(t)\pi(\boldsymbol\theta', \boldsymbol{D}', K+1)q_d(\boldsymbol{u}').
\end{equation}
In similar fashion to reversible jump MCMC \citep{Green1995} and birth-death MCMC \citep{Stephens2000}, we also require that the transformation that maps $(\boldsymbol\theta,\boldsymbol{u}) \mapsto (\boldsymbol\theta', \boldsymbol{u}')$ is a bijection and that a dimension matching condition is met. To satisfy these conditions birth moves are defined by drawing parameters for a new hazard, $\boldsymbol{u}$, from the prior conditional on $\boldsymbol\phi$ and selecting the distribution of the new hazard uniformly at random. The reverse move then selects a hazard uniformly at random to remove from the model.

To satisfy \eqref{eq;BD_balance}, a simple way of specifying $\Lambda^b(t)$ is via a balancing function \citep[e.g.~][]{Zanella2019} $b : \mathbb{R}_+ \to \mathbb{R}_+$ satisfying $b(a) = a\cdot b(1/a)$, and taking the Metropolis--Hastings--Green ratio
\begin{equation*}
    a(t) := \frac{\pi(\boldsymbol\theta'_t, \boldsymbol{v}'_t, \boldsymbol\phi, \boldsymbol{D}', K+1)q_D(\boldsymbol{u}')}{\pi(\boldsymbol\theta_t, \boldsymbol{v}_t, \boldsymbol\phi, \boldsymbol{D}, K)q_B(\boldsymbol{u})},
\end{equation*}
as its argument. The required death move is then defined similarly but with the argument $a(t)^{-1}$. The most commonly used example of this is the Metropolis balancing function, $b_M(a) = \min\{1,a\}$, which is the foundation of the Metropolis-Hastings algorithm. Extended theoretical justification of this approach and further discussion of the role of balancing functions is provided in the supplementary materials.

An alternative specification, which is the birth-death MCMC approach, is to take $\Lambda^b(t)$ constant and set $\Lambda^d(t) = a^{-1}$. This method fails in our setting as, in contrast to \cite{Stephens2000}, $\theta$ is being updated in continuous-time. The resulting ratio of posterior densities is then challenging to upper bound, which is needed to apply Poisson thinning. Note, however, that $b_M(a) \leq 1$ and therefore this birth rate is amenable to Poisson thinning. The specification of $b_M(a)$ holds up to a multiplicative constant, $\Lambda^{K}$, which can be used to control the intensity of transdimensional updates.

\subsubsection{Swap moves}
\label{sec;Swap}
While the birth-death process is sufficient to sample from the correct target distribution, we find that posterior exploration can be significantly improved by the introduction of moves which swap subhazard distributions without updating $K$. These allow the sampler to move between models with the same number of hazards but different underlying distributions. The improvement in mixing is most noticeable when the posterior for $K$ is concentrated but the posterior for $\boldsymbol{D}\mid K$ is more diffuse, as it avoids the need for transitions through higher or lower order hazard models with low posterior mass.

We define our swap moves between distributions based on the principle of median matching. Moment matching is a well established approach in defining reversible jump moves \citep{Richardson1997} but is not applicable here as moments for some survival distributions are not well defined (e.g the log-logistic distribution with $\nu < 1$). 

Median matching is a deterministic proposal in which the distribution of a subhazard is swapped from log-logistic to Weibull or vice versa. Considering the case without covariates first, the method keeps the shape parameters of the old and new hazards the same, and then transforms the location parameter to keep the medians the same, using the formula
\begin{gather*}
    \text{Med}_{LL}(\nu,\mu) = \mu = \left(\frac{1}{\mu'}\right)^{\frac{1}{\nu}}(\log2)^{1/\nu} = \text{Med}_{W}(\nu,\mu'), \\
    \implies \mu' = \mu^{-\nu}\log2.
\end{gather*}

When including standardised covariates, the interpretation of the above is that the subhazard median is preserved for the average individual. To include covariates in the transformation we apply the mapping $\beta_{LL} \mapsto -\beta_{LL} = \beta_W$. Intuitively it seems reasonable to expect the magnitude of the coefficient effects to be the same when altering the subhazard distribution. However, the interpretation of the effect is inverted, hence the switching of the sign. The median matching proposal can be placed into the balancing function framework outlined previously, although the Metropolis--Hastings--Green ratio now requires a Jacobian to account for the transformation.

\begin{figure}
    \centering
    \includegraphics[width=5.5in]{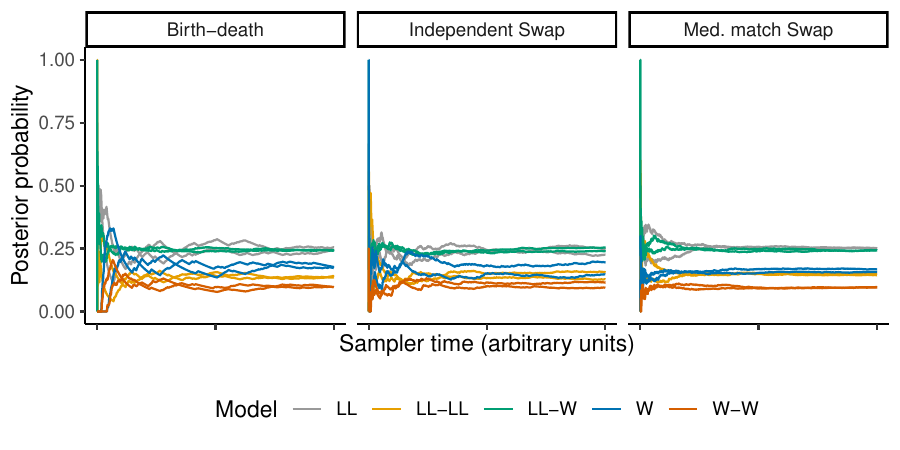}
    \caption{Experiment comparing the efficiency of the median matching swap moves to the birth-death moves and independent swap moves, on data simulated from a poly-log-normal-Weibull model with a single covariate. Coloured lines represent different subhazard combinations. Two chains for each method were produced running for 10,000 time units, with reversible jump moves occurring at the same rate. The median matching swap moves provide more stable and efficient mixing in comparison to the alternative methods.} 
    \label{fig:Swap}
\end{figure}

Figure \ref{fig:Swap} shows trace plots of posterior model probabilities for samplers using solely the birth-death process; independent swaps; and median matching swaps; based on data containing 100 simulated survival times and a single binary covariate. Note that swap moves and birth-death moves have the same computational cost and, as the overall birth-death-swap rate was set to 10 in each case, the expected computational cost is identical for each sampler. Almost all the posterior mass is placed on models such that $K < 3$, but posterior mass is spread relatively evenly between these models. The median matching moves provide clearly superior convergence in comparison to the alternative processes, where slow convergence is observed for the log-logistic and Weibull models as posterior exploration between these models requires moving through higher order models. Acceptance rates for independent swaps and median match swaps were respectively 6.09\% and 44.17\%, showing the clear superiority of the bespoke moves.  

\subsection{Practical implementation}
\label{sec;Impl}
The methodology outlined in this Section requires the generation of multiple event times simultaneously. For computational efficiency this is done via the multinomial trick, whereby a single event time is generated with rate equal to the sum of rates and then a single event is chosen with probability proportional to its rate. Times until deterministic sticking events are also simply tracked and updated when necessary. For clarity we provide the full summary of the process in the supplementary material.

\subsubsection{MCMC output}
\label{sec;Output}
As stated previously the Zig-Zag sampler outputs piecewise continuous sample paths. This can be stored either as a skeleton of points which indicate updates to one of $(\boldsymbol{v},K,\boldsymbol{D},\boldsymbol\gamma)$, or as samples at exponential times. The effect of this and the rate of drawing samples is analogous to the role of thinning in discrete time MCMC.

For identifiability purposes we place an ordering constraint on the shape parameters of hazards with the same distribution as a post-processing step to sort the MCMC output. This is appropriate in this setting as: a) The quantity of interest, mean survival, is not invariant to permutation, and so our inference should not suffer due to the re-labelling issue. b) \cite{Kozumi2004} explored the use of loss functions in the poly-Weibull model and found that the resulting inferences where almost identical to the use of an ordering constraint. We therefore believe that alternative approaches would have little benefit, and that the ordering constraint is sufficient when examining individual subhazards during, for example, model checking.

To summarise, our approach utilises the sticky Zig-Zag sampler to update $\boldsymbol\theta, \boldsymbol\gamma \mid K, \boldsymbol{D}, \boldsymbol\phi$ in continuous-time using gradient information and non-reversible dynamics to ensure efficient exploration of the posterior. This sampler is combined with continuous-time jump processes for updating $K, \boldsymbol{D}, \boldsymbol\phi$ based on conjugate updates, adaptive Metropolis steps and a bespoke birth-death-swap process. The shared continuous-time framework allows for events to be efficiently generated via Poisson thinning and the multinomial trick.

\section{Real data case studies}
In this Section we apply the methodological extensions to polyhazard models proposed in the previous two Sections to three real world examples focusing on the effect of prior specification on computation and inference and the non-linear covariate effects produced by polyhazard models. 

\label{sec;Applications}
\subsection{Lung transplant data}
\label{sec;Lung}
\cite{Demiris2015} used poly-Weibull models to calculate mean survival in lung transplant patients, focusing particularly on differences between patients who received single and double lung transplants. The data contain survival or censoring times of 338 patients, 173 (144 observed) of whom received single lung transplants and 165 (79 observed) of whom received double lung transplants. They focus their analysis on a set of `highly likely` variations of the poly-Weibull model, as assessed by the mean deviance, all of which indicate small differences in early survival but higher risk for single lung transplant patients in the long-term. This is due to a partial treatment effect, which increases the risk patients experience over a lifetime time horizon.

Although the original data are not publicly available we have constructed a similar dataset by digitising Figure 1 of \cite{Demiris2015}. This was done using the implementation of the method of \cite{Guyot2012} available via the \texttt{Survhe} \texttt{R} package. We re-analyse these data with the same objective using the extended polyhazard model. We set $\sigma_\alpha = 2$, $K_{\max} = 4$, and adjust the above prior structure by fixing $\sigma_\beta = 5$ and $\omega = 0.5$, which prevents $(\omega, \sigma_\beta)$ from being essentially nonidentifiable in the presence of a single covariate. 

The number of candidate models in this scenario is 128, which, although possible to explore manually, would still be computationally expensive. Our approach has the dual advantage of saving computational cost by focusing on models with high posterior probability, and also providing posterior probabilities for each sub-model.

The samplers were run for 10,000 time units with a sampling rate of 4 to generate approximately 4 samples per time unit, and reversible jump rate of 10. With birth, death and swap acceptance rates of 4.90\%, 4.89\% and 6.10\% respectively this results in an across model update approximately every 2 time units. The sampler took on average 10 minutes to run. Sampler trace plots and diagnostics are available in the supplementary materials.

\begin{table}[]
    \centering
    \begin{tabular}{c|c|c|c|c}
        Model & Post. probability & Mean survival DLT & Mean survival SLT & Difference \\
        \hline
        Averaged model & --- & 7.58 (5.44, 13.05) & 4.50 (3.73, 5.44) &  3.08 (0.84, 8.61) \\
        Original W-W & --- & 8.78 (6.14, 13.7) & 4.96 (4.32, 5.75) & 3.83 (1.04, 8.72) \\
        W-L & 0.202 & 7.58 (5.42, 12.20) & 4.42 (3.67, 5.3) &  3.16 (0.91, 7.76) \\
        W-W & 0.011 & 8.18 (5.56, 11.75) & 4.64 (3.69, 6.24) & 3.54 (0.80, 6.99)\\
        L-L & 0.624 & 7.57 (5.45, 13.35) & 4.52 (3.75, 5.44) & 3.05 (0.82, 8.86) \\
        W-W-L & 0.018 & 7.59 (5.41, 11.66) & 4.42 (3.72, 5.27)  & 3.16 (0.87, 7.21) \\
        W-L-L & 0.065 & 7.55 (5.44, 12.79) & 4.47 (3.72, 5.4) & 3.08 (0.82, 8.39)  \\
        L-L-L & 0.066 & 7.52 (5.42, 13.04) & 4.49 (3.73, 5.37) & 3.02 (0.80, 8.69) \\
    \end{tabular}
    \caption{Model summaries for the averaged model, original model, and sub-models with $>1\%$ posterior mass. Mean survival estimates are shown for single (SLT) and double (DLT) lung transplant patients along with the expected difference in survival (and relevant 95\% credible intervals). Estimates from the original bi-Weibull model are as reported in the original analysis.}
    \label{tab:DemTable}
\end{table}

Table \ref{tab:DemTable} shows model summaries for the original bi-Weibull model chosen by \cite[][original W-W]{Demiris2015}, the averaged polyhazard model and all submodels with posterior probability greater than 1\%. Notably the original bi-Weibull model receives less than 2\% posterior probability, with the majority of the posterior mass  focused on the bi-log-logistic model (62.5\%), with reasonable mass on the Weibull-log-logistic model (19.9\%) and 15.5\% posterior probability shared between three of the three hazard models.

Mean survival estimates for single (SLT) and double (DLT) lung transplant patients are more conservative than those reported in the original analysis. As the reduction in DLT survival is larger than for SLT survival, the analysis using the averaged model reports a smaller difference in expected survival. Although this disparity is driven by a preference for the bi-log-logistic model, the estimates from the bi-Weibull sub-model also suggest more conservative survival estimates and a smaller difference in survival. These differences are discussed in Section \ref{sec;Prior_sens}. Negligible posterior mass was placed on the single hazard models, corroborating the results from the original analysis, which suggested that single hazard models were insufficient.

\begin{figure}
    \centering
    \includegraphics[width=5.5in]{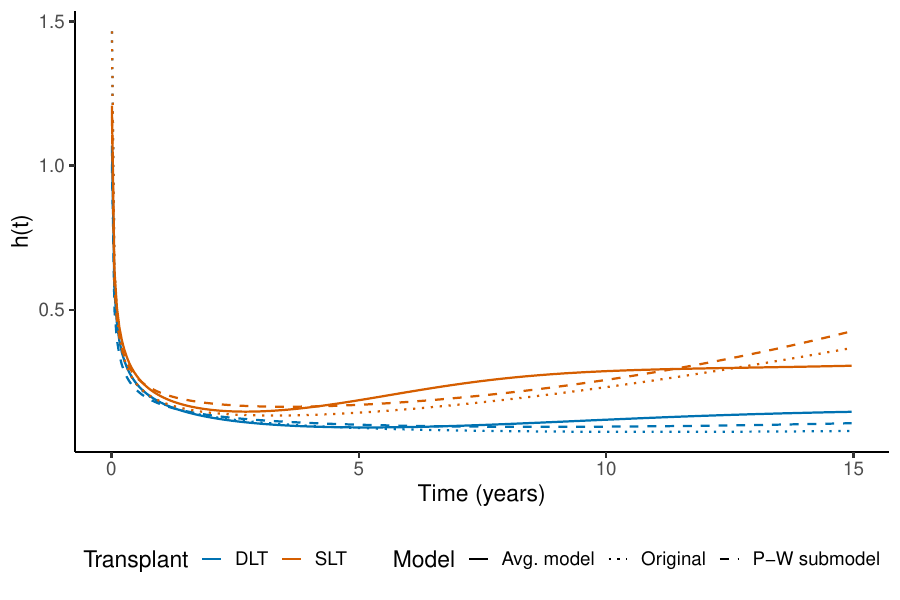}
    \caption{Hazards for different models fit to the lung transplant data. The hazards for the model from the original analysis (dash-dot), from the bi-Weibull model in our analysis (dashed) and from the overall hazard from our analysis (solid). These are plotted for DLT (blue) and SLT (red) patients.}
    \label{fig:Dem_haz}
\end{figure}

Figure \ref{fig:Dem_haz} shows hazards for SLT and DLT patients from the overall model, the bi-Weibull model from our analysis and the bi-Weibull model from the original analysis. Notably all three models produce very similar results in the short-term and only differ noticeably after 3 years. This suggests the difference in results reported in Table \ref{tab:DemTable} is due to differences in hazards for long-term survivors. Compared to the original analysis the hazard for SLT patients increases faster than in the original analysis after five years explaining the difference in the results reported in Table \ref{tab:DemTable}.

A key foundation of the original analysis is that the bathtub curve is commonly observed for transplant patients. This can be seen in our example where, although the bi-log-logistic model with highest posterior probability is not a bathtub curve, a decreasing-increasing pattern is observed over a typical patient lifetime, with the overall mixture of polyhazard models ensuring that as $t \to \infty$ we observe $h(t) \to \infty$.

\subsubsection{Weakly informative priors}
\label{sec;Prior_sens}

The two bi-Weibull models in Table \ref{tab:DemTable} report different estimates of difference in mean survival between transplant types. While some of this difference arises from the data digitisation process, this is also due to the use of weakly informative rather than non-informative prior information. 

Increasing the standard deviation of the prior for $\boldsymbol\beta$, increases the posterior estimate for mean survival in both arms and the corresponding credible intervals. This is due to the increasing mass placed on extreme mean survival values by the increasingly non-informative prior. In a single hazard model this is not problematic as the likelihood provides sufficient regularisation of $\beta_{10}$. In a $K$ hazard model, however, this behaviour results in the $k^{th}$ subhazard having negligible influence on the likelihood and the model in effect reducing to a $K-1$ hazard model. This has the combined effect of hindering computation, whether via Gibbs sampling or using gradient-based samplers, and impairing the resulting inference. We note that this effect is independent of the prior for $\gamma$ which as historically been the focus of identifiability in polyhazard models.

As the standard deviation of the priors for $\boldsymbol\beta$ increase so do the posterior estimates along with the size of their credible intervals due to the increasing mass placed on extreme mean survival values by the increasingly non-informative prior. In a single hazard model this is not problematic as the likelihood provides sufficient regularisation of $\beta_{10}$. In a $K$ hazard model, however, this behaviour results in the $k^{th}$ subhazard having negligible influence on the likelihood and the model in effect reducing to a $K-1$ hazard model. This has the combined effect of hindering computation, whether via Gibbs sampling or using gradient-based samplers, and impairing the resulting inference. We note that this effect is independent of the prior for $\gamma$ which as historically been the focus of identifiability in polyhazard models.

This undesirable behaviour can be excluded by the use of weakly informative priors for $\beta_{k,0}$, as outlined in Section \ref{sec;Priors}. Although tighter than those used previously in the literature, we would argue that these priors are still weakly informative in that they are able to generate data and inferences well beyond the range of plausible values following similar arguments made in \cite{Gabry2019}. We recommend conducting prior sensitivity analysis to ensure this regularisation is sufficient but not unnecessarily influential. In cases with a large number of candidate models, this can be focused on the small subset of models with high posterior probability to preserve computational efficiency.

\subsection{COST data}
\label{sec;COST}
We now apply the methodology to a more challenging example -- data from the Copenhagen Stroke Study (COST), a prospective, cohort study of stroke survivors in Copenhagen starting in 1991 \citep{Jørgensen1996}. In this setting, standard parametric models are likely to be insufficient as we may expect the hazard to evolve from reflecting the short-term risks immediately post-stroke to the longer-term risks stroke survivors face and the increase in risk with increasing age. Further, given the high number of covariates in the dataset, the standard approach of fitting separate models to subgroups for extrapolation is likely to be insufficient due to the increase in uncertainty associated with smaller sample sizes.

A subset of the data containing survival times, event indicators and 13 covariates, including those discussed previously, for 518 patients is available via the \texttt{pec R} package \citep{Mogensen2012}. A complete summary of the dataset is provided in the supplementary materials.

\subsubsection{COST results}
\label{sec;COST_results}
We fit the model using the full prior structure outlined in Section \ref{sec;Priors}. The sampler was run for 20,000 time units, with the jump rate set to 10. At each jump a birth-death move, swap move or hyperparameter update was attempted each with probability $1/3$. The sampler took approximately 90 minutes to run per chain. Trace plots, diagnostics and prior sensitivity checks are provided in the supplementary materials. Birth, death and swap acceptance rates were 4.36\%, 4.32\% and 1.99\% respectively.

Given the larger sample size and lower censoring rate posterior submodel probabilities are relatively concentrated, with 86.55\% of the posterior mass given to the bi-log-logistic model, 6.48\% to the tri-log-logistic model, 4.90\% to the W-L-L model, and 1.67\% to the Weibull-log-logistic model. All other models have less than 1\% posterior mass.

\begin{figure}
    \centering
    \includegraphics[width=5.5in]{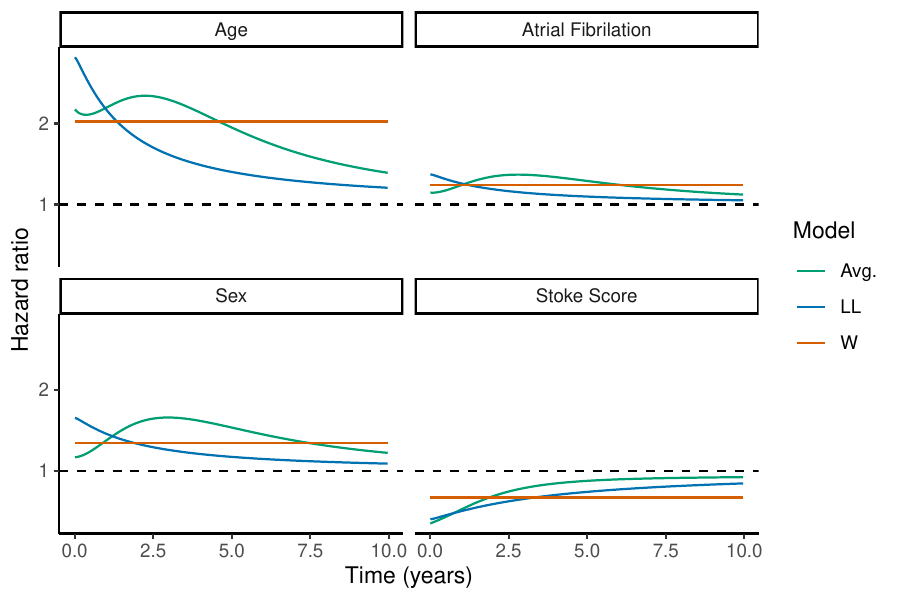}
    \caption{Hazard ratios (HRs) for atrial fibrilation, age, sex and stroke score from the COST dataset. The green line is the HR from the averaged polyhazard model (Avg.), the blue and orange lines are hazard ratios obtained from the simpler log-logistic (LL) and Weibull (W) models. A hazard ratio of 1 is indicated by a dashed line on each plot.}
    \label{fig:COST_HRs}
\end{figure}

An advantage of using polyhazard models is the ability to model covariate effects more flexibly than under standard assumptions of proportional hazards or accelerated failure times. This can be seen in Figure \ref{fig:COST_HRs}, where we plot the hazard ratios over time for atrial fibrillation, age, sex and stroke score and compare them to the hazard ratios for the simpler Weibull and log-logistic models. For continuous covariates these are defined as the hazard ratio between the observed 25\% quantile and 75\% quantile in the data. Notably the averaged model is able to capture a wide variety of flexible hazard ratios. 

The hazard ratio for age suggests older stroke sufferers have a higher risk of death, which decreases but remains notable for 10 years post-stroke. This aligns with the analysis of \cite{Kammersgaard2004}. The hazard ratio for sex corroborates the findings of \cite{Andersen2005} that women have higher survival than men, although it suggests that the difference in risk decreases in time after an initial peak.  A similar pattern is observed for atrial fibrillation. Stroke severity (as measured by stroke score), shows that survivors of less severe strokes are at lower risk of death in the short-term, but that this difference in risk becomes less prevalent in the long-term. In each case the single Weibull and log-logistic hazard ratios are unable to match the increased flexibility of the polyhazard model.

\begin{figure}
    \centering
    \includegraphics[width=5.5in]{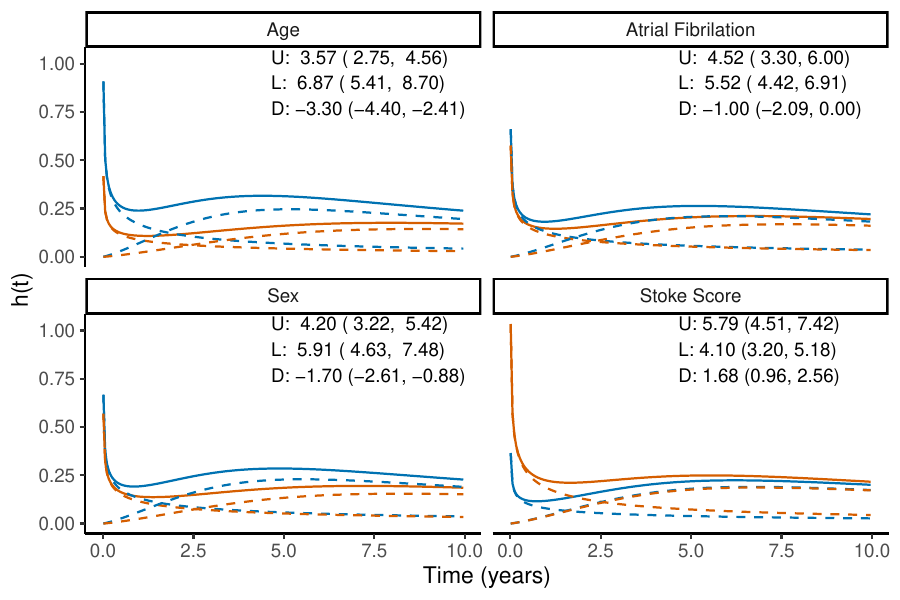}
    \caption{Hazards for different values of atrial fibrilation, age, sex and stroke score from the COST dataset. Arial Fibrilation and sex: 0 (orange), 1 (blue). Age and Stroke Score: Sample lower quartile (orange), sample upper quartile (blue). Other covariates are set to 0 representing the average patient. Boxes in each plot show mean survival estimates for the blue hazard (U), orange hazard (L) and the difference in mean survival (D).}
    \label{fig:COST_haz}
\end{figure}

Figure \ref{fig:COST_haz} plots hazards for each covariate group from the overall models (solid lines) and from the two hazards from the dominating bi-log-logistic model (dashed lines) for the same covariates. Interpreting the first hazard as the immediate post-stroke risk and the second as the longer-term risks, we can understand the influence of different covariates. In particular age increases both the immediate risk post-stroke and the long-term risk, while atrial fibrillation and being male has no immediate effect, but a noticeable long-term effect. Conversely, less severe strokes reduce risk in the short-term but have a less noticeable effect in the long-term. Figure \ref{fig:COST_haz} also contains estimates of mean survival and difference in mean survival. In each of the highlighted covariates the 95\% credible interval for difference in mean survival does not contain 0, although for atrial fibrillation it coincides with the boundary of the interval, presenting clear evidence that the presence of atrial fibrillation, increasing age and being male lower survival, while less severe strokes improve survival.


\subsection{Taiwan Kidney Transplant data}
\label{sec;TKT}
We apply our methodology to data on survival times of 3,562 Taiwanese patients following uncomplicated kidney transplantation with the primary objective of understanding the impact of waiting times on mean survival \citep{Chen2022a}. The data were accessed via Dryad \citep{Chen2022b}. 

Using the prior structure in Section \ref{sec;Priors} we fit the averaged model to this data. We make the modification of only considering models with $K < 4$ as, given the high censoring rates, it is unlikely that there is sufficient information in the data to define more than 3 hazards. Further we use a slightly more informative $\text{Normal}(0,1)$ prior for the shape parameters as we otherwise encounter identifiability issues similar to those highlighted in Section \ref{sec;Prior_sens}. The sampler was run for 20,000 time units, with the rate of reversible jumps or Gibbs moves set to 20. This took 11.76 hours to run.

\begin{table}[]
    \centering
    \begin{tabular}{c|ccccccc}
        Model  & W-L & W-W & L-L & W-W-L &  W-L-L & W-W-W &  L-L-L  \\
        \hline 
        Post. prob. & 0.212 & 0.416 & 0.015 & 0.164 & 0.090 & 0.092 & 0.008 \\
        \hline
    \end{tabular}
    \caption{Posterior sub-model probabilities for the averaged model applied to the Taiwanese Kidney Transplant dataset restricted to models with posterior mass above $0.005$.}
    \label{tab:TKTTable}
\end{table}

Posterior model probabilities are reported in Table \ref{tab:TKTTable}. The majority of the posterior mass is shared between the bi-Weibull, Weibull-log-logistic and bi-Weibull-log-logistic models. The posterior is less concentrated than in the previous examples, due to the limited complete data in the sample. 

\begin{figure}
    \centering
    \includegraphics[width=5.5in]{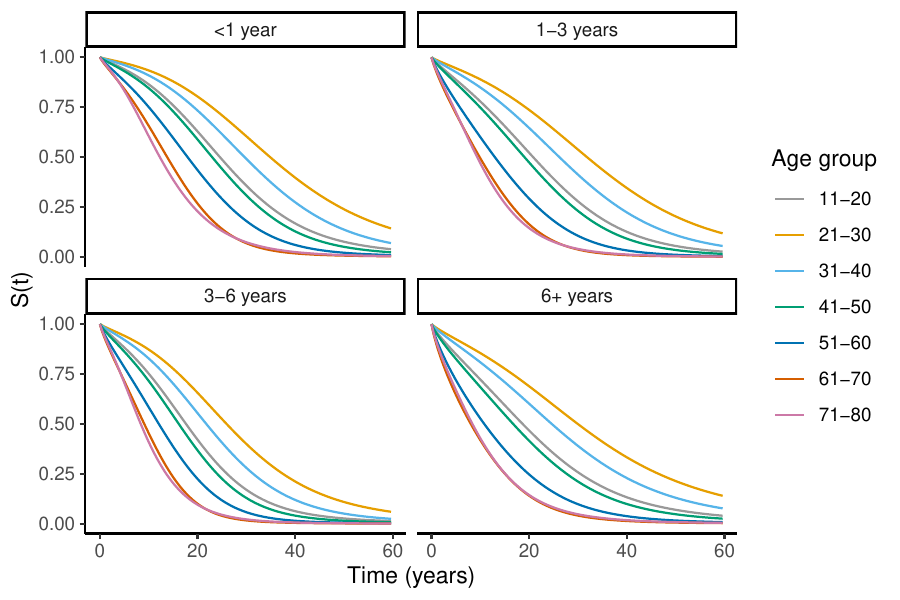}
    \caption{Mean survival curves from the averaged model for the Kidney transplant data set stratified by waiting time and age.}
    \label{fig:TKT_surv}
\end{figure}

Figure \ref{fig:TKT_surv} shows survival curves for each waiting time group stratified by age. Each curve appears to reach 0 in a reasonable time frame. Of particular note is the apparently non-linear effect of age, with patients in the youngest age group (11-20) having worse survival than patients aged 21-40. This effect is not implausible due to the differing reasons for requiring a kidney transplant in different age groups, which are possibly more likely to be due to genetic or hereditary conditions for younger patients, and more likely due to lifestyle factors in older patients. Further in all waiting time groups there are minimal differences in survival between patients in the oldest age groups. 

\begin{figure}
    \centering
    \includegraphics[width=5.5in]{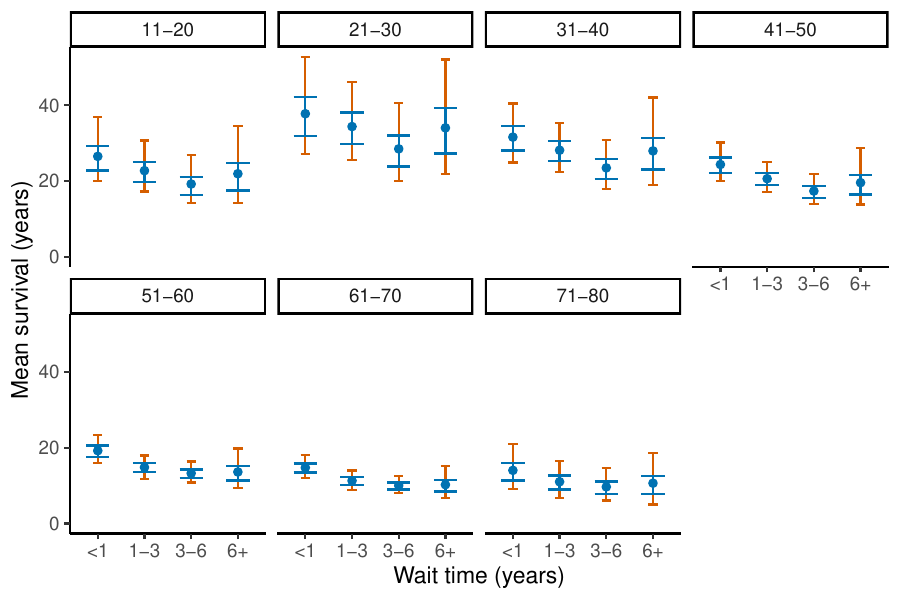}
    \caption{Posterior summaries for mean survival, with posterior mean (solid blue dot), 50\% credible interval (blue, larger, error bar) and 95\% credible intervals (orange, smaller error bar). Results are stratified by age and waiting time.}
    \label{fig:Wait_surv}
\end{figure}

\begin{figure}
    \centering
    \includegraphics[width=5.5in]{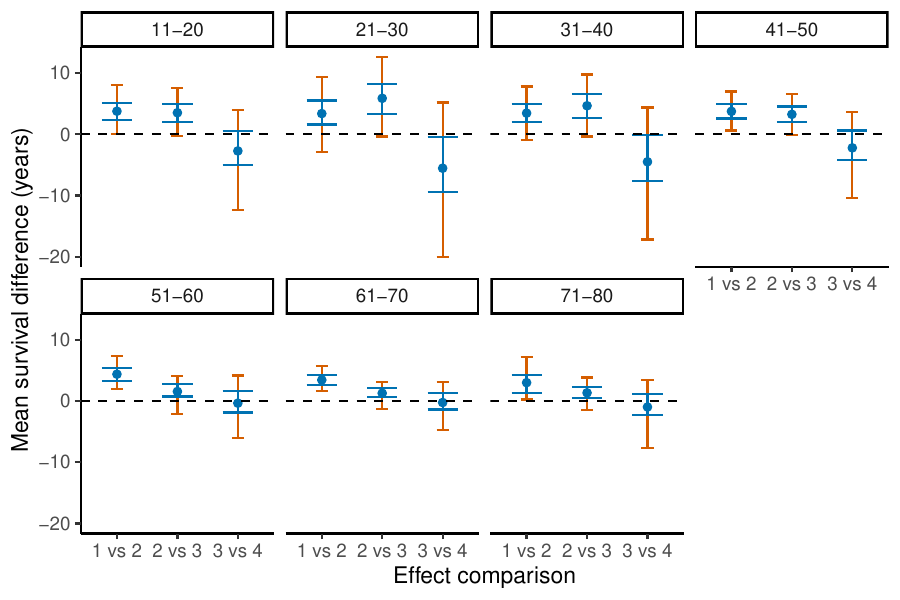}
    \caption{Posterior summaries for mean survival difference, with posterior mean (solid blue dot), 50\% credible interval (blue, larger, error bar) and 95\% credible intervals (orange, smaller error bar). Results are stratified by age and waiting times encoded as (1: <1 years, 2: 1-3 years, 3: 3-6 years, 4: 6+ years). A dashed line is used to indicate 0 difference.}
    \label{fig:Wait_diff}
\end{figure}

To understand the effect of waiting times on mean survival, posterior estimates of mean survival stratified by age and waiting time group are presented in Figure \ref{fig:Wait_surv}, with posterior means, 75\% and 95\% credible intervals plotted. Similarly, the effect of moving reducing waiting time by one group is shown in Figure \ref{fig:Wait_diff}. The uncertainty associated with these estimates reduces with age in both cases as the number of censored observations decreases, except for the oldest age group which corresponds to only 17 patients in the sample, resulting in very high uncertainty. Similarly there is high uncertainty in each age group for mean survival in patients who waited more than 6 years for a transplant which propagates through to the estimates of difference in mean survival between patients who waited 6+ years and those who waited 3-6 years. From Figure \ref{fig:Wait_diff} there is strong evidence to suggest that in the youngest age group and patients over 51 reducing waiting times from 1-3 years to $<$1 year improves mean survival and similarly reducing wait times from 3-6 years to 1-3 years for patients under 50 improves mean survival. In each age group the lack of information for patients with wait times over 6 years means there is high uncertainty related to the corresponding effect size.

\section{Discussion}
\label{sec;Discussion}

In this work we have developed an extended version of the polyhazard model, using an extended prior specification and novel posterior sampling methodology. This allows for the efficient application of polyhazard models to two motivating data sets for which previous approaches to model selection and computation would have been infeasible. Further, through the use of Bayesian model averaging, we limit the risk of survival extrapolation and mean survival inferences being affected by model misspecification when compared to selecting a single best model.

The findings from the analysis of the digitised lung transplant data from \cite{Demiris2015} suggest that non-informative priors are not appropriate in the polyhazard model setting as they place too much mass on unreasonably large mean survival values. This results in poor posterior estimates and identifiability issues not previously commented on in the literature.

The analysis of the COST dataset shows how the polyhazard model is able to translate epidemiological findings to a cost-effectiveness analysis in the presence of covariates. In particular our approach circumvents issues with current approaches, that either fit models for each subgroup or rely on strong covariate assumptions. The analysis of the kidney transplant data set shows that the extended polyhazard model is able to account for high censoring rates. In particular, being able to combine estimates from many plausible models provides more principled extrapolations in the presence of partial information. 

The approach of this paper is an addition to a number of methods which seek to provide more principled extrapolations by learning the parameters for extrapolation primarily from data towards the end of the observation period. Other examples include the use of M-splines \citep{Jackson2023} and dynamic survival models \citep{Kearns2022}. In contrast to these methods, our approach retains an increased degree of interpretability but a more thorough comparison of these approaches would be beneficial.

We note that the extended polyhazard model can be easily combined with several methods for improving extrapolations and integrating external information. In particular polyhazard models are the natural form for integrating external information, whether this relates to specific causes of death \citep{Benaglia2015} or life table data for the wider population \citep{vanOostrum2021}. Alternatively, the extended polyhazard model could be used to model the observed period and then combined with life-table data via the blended survival approach of \cite{Che2023}. Further simple adjustments to the model could also be made to combine it with other model averaging approaches to extrapolation. For example, the adjusted model averaging approach of \cite{Negrin2017} can be combined with our methods by adjusting posterior weights to account for optimistic and skeptical scenarios.

We briefly outline some obvious extensions to the model presented in \ref{sec;Priors}. Firstly, we can naturally extend the model to include additional subhazard forms. Although there are many two-parameter survival distributions in the literature, selecting a small number of additional distributions should provide sufficient flexibility to model many datasets. In this context the swap moves from Section \ref{sec;MCMC} could be extended to define pairwise transformations between different types of subhazards, or replaced with moment-matching moves where appropriate. Another novel extension would be to introduce the possibility of improper subhazards such that for the corresponding survivor functions 
\begin{equation*}
    S_{k,\theta}(t) \to c > 0, \quad t \to \infty.
\end{equation*}
This would correspond to a cure model for that subhazard, but would need highly informative external information to ensure principled extrapolations. A final extension would be to introduce dependence between hazards, as explored by \cite{Tsai2013}.

Finally, we believe we have made important contributions to the applications of PDMP samplers. While these samplers have seen several methodological and theoretical developments, they have seen limited practical application. We hope that their usage in this work can motivate their usage in other contexts. In particular, the bounding method developed in Section \ref{sec;MCMC} is not model dependent so could be applied in other contexts, as could the extension of the Gibbs Zig-Zag approach to transdimensional updates. In the context of PDMP samplers for variable selection, the combination of variable selection dynamics with the Gibbs Zig-Zag approach for updating hyperparameters efficiently is an important advancement, which can avoid the use of fixed spike and slab weights. Finally, the median matching heuristic developed for the swap moves may be useful in other contexts.

%%%%%%%%%%%%%%%%%%%%%%%%%%%%%%%%%%%%%%%%%%%%%%
%% Support information, if any,             %%
%% should be provided in the                %%
%% Acknowledgements section.                %%
%%%%%%%%%%%%%%%%%%%%%%%%%%%%%%%%%%%%%%%%%%%%%%
\begin{acks}[Acknowledgments]
%We would like to thank Nikos Demiris for access to the lung transplant data and permission to use it in this paper as well as constructive comments on an earlier version of this paper.
\end{acks}

%%%%%%%%%%%%%%%%%%%%%%%%%%%%%%%%%%%%%%%%%%%%%%
%% Funding information, if any,             %%
%% should be provided in the                %%
%% funding section.                         %%
%%%%%%%%%%%%%%%%%%%%%%%%%%%%%%%%%%%%%%%%%%%%%%
\begin{funding}
LH is supported by EPSRC grant EP/W523835/1 and the Alan Turing Institute Enrichment Scheme. SL is partially supported by EPSRC grant EP/V055380/1.
\end{funding}

%%%%%%%%%%%%%%%%%%%%%%%%%%%%%%%%%%%%%%%%%%%%%%
%% Supplementary Material, including data   %%
%% sets and code, should be provided in     %%
%% {supplement} environment with title      %%
%% and short description. It cannot be      %%
%% available exclusively as external link.  %%
%% All Supplementary Material must be       %%
%% available to the reader on Project       %%
%% Euclid with the published article.       %%
%%%%%%%%%%%%%%%%%%%%%%%%%%%%%%%%%%%%%%%%%%%%%%
\begin{supplement}
\stitle{Supplementary material: Further sampling details and experiments, and modelling details}
\sdescription{Provides a summary of the sampling process described in Section \ref{sec;MCMC}; theoretical justification of the extension of the Gibbs Zig-Zag method to transdimensional updates; details of the experiment for the efficacy of the median-matched swap moves; trace plots and summaries of the models fit in Section 4. The methods discussed in this work were implemented in Julia. Full code to implement these models and re-create our results, and the digitised lung transplant data is available at the GitHub repository: \url{https://github.com/LkHardcastle/PolyhazardPaper}.}
\end{supplement}
%%  imsart-nameyear.bst  will be used to                   %%
%%  MR numbers will be added by VTeX.                      %%
%%                                                         %%
%%  Use \cite{...} to cite references in text.             %%
%%                                                         %%
%%%%%%%%%%%%%%%%%%%%%%%%%%%%%%%%%%%%%%%%%%%%%%%%%%%%%%%%%%%%%

%% if your bibliography is in bibtex format, uncomment commands:
%\bibliographystyle{imsart-nameyear} % Style BST file
%\bibliography{bibliography}       % Bibliography file (usually '*.bib')

%% or include bibliography directly:

\end{document}

\begin{frontmatter}
\title{Supplementary material A: Further sampling details and experiments}
%\title{A sample article title with some additional note\thanksref{T1}}
\runtitle{Supplementary materials A}
%\thankstext{T1}{A sample of additional note to the title.}

\begin{aug}
%%%%%%%%%%%%%%%%%%%%%%%%%%%%%%%%%%%%%%%%%%%%%%%
%% Only one address is permitted per author. %%
%% Only division, organization and e-mail is %%
%% included in the address.                  %%
%% Additional information can be included in %%
\author[A]{\fnms{Luke}~\snm{Hardcastle}\ead[label=e1]{luke.hardcastle.20@ucl.ac.uk}},
\author[A]{\fnms{Samuel}~\snm{Livingstone}\ead[label=e2]{samuel.livingstone@ucl.ac.uk}}
\and
\author[A]{\fnms{Gianluca}~\snm{Baio}\ead[label=e3]{g.baio@ucl.ac.uk}}
%%%%%%%%%%%%%%%%%%%%%%%%%%%%%%%%%%%%%%%%%%%%%%
%% Addresses                                %%
%%%%%%%%%%%%%%%%%%%%%%%%%%%%%%%%%%%%%%%%%%%%%%
\address[A]{Department of Statistical Science,
University College London\printead[presep={,\ }]{e1,e2,e3}}

\end{aug}

\end{frontmatter}

\section{The sampler}

We present a concise summary of the sampler outlined in Section 3. Given $(\theta, v, \phi, \gamma, D, K)$ at sampler time $t$,  the process updates $\theta$ as 
\begin{equation*}
    \theta_{t+s} = \theta_t + v_ts,
\end{equation*}
until one of the following occurs:
\begin{enumerate}
    \item An update to $v$ with rate given by $\Lambda^F(t) = \sum_{i=1}^d\Lambda^F_i(t)$, where $\Lambda^F_i(t)$ is the rate associated with the $i^{th}$ coordinate. These events are generated via Poisson thinning using the bounding rate $M(t) + \Lambda_0$, where $M(t)$ is found using the method outlined in Section 3.1.1 and $\Lambda_0$ is an offset accounting for numerical errors in the bounding process.
    \item An update to $(\gamma,v)$ which occurs
    \begin{enumerate}
        \item with rate $(1-\omega)/\omega$ when $\gamma_{k,j} =0$.
        \item when $\beta_{k,j} = 0$ for $\gamma_{k,j} = 1$.
    \end{enumerate}
    \item An update to $\phi$ via Gibbs/Metropolis-within-Gibbs steps which occur with constant rate $\Lambda^H$.
    \item An update to $(\theta,D)$ via a swap step. These occur with rate $\Lambda^S(t)$, generated via Poisson thinning with the upper bound $\Lambda^S$.
    \item An update to $(\theta, v, \gamma, D, K)$ via a birth or death move. These occur with rates $\Lambda^B(t)$ and $\Lambda^D(t)$ respectively and are generated using Poisson thinning using the upper bound $\Lambda^{BD}$.
\end{enumerate}
To use the multinomial trick the first time from the Poisson process with rate
\begin{equation*}
    M(t) + \Lambda_0 + \frac{1-\omega}{\omega}\sum_{(k,j)}1(\gamma_{k,j}=0) + \Lambda^H + \Lambda^S + \Lambda^{BD},
\end{equation*}
is generated, and then a type of move is selected with probability proportional to the corresponding rate. In the case of the birth-death-swap process and velocity flips a Poisson thinning step then needs to be carried out.
\section{Birth-death-swap MCMC within Zig-Zag sampling}

We present an argument for the validity of the transdimensional moves by extending the arguments presented in \cite{Sachs2023}. The main idea is to replace the Gibbs kernel in (5) of \cite{Sachs2023} with a reversible jump kernel \citep{Green1995}, including a Jacobian to account for the corresponding transformation (as is the case with the median-matching swap moves). 

Without the underlying PDMP sampler this would correspond to birth-death MCMC \citep{Stephens2000}, although with an alternative specification of jumping rates, and would (inefficiently) provide valid posterior samples. Following arguments from \cite{Sachs2023}, we can then superimpose the generators for this process, the Zig-Zag sampler for sampling $(\theta, \gamma)$ and the (Metropolis-within-)Gibbs updates for hyperparameters to construct a process with the correct target distribution.

\subsection{On the role of balancing functions}

In this work we use the Metropolis balancing function to define the birth-death-swap process for updating $(K,D)$. Other choices of balancing function are available, however, for example the barker balancing function,
\begin{equation*}
    b_B(a) = \frac{a}{1+a}.
\end{equation*}
In the context of discrete time MCMC \cite{Peskun1972} showed that the Metropolis balancing function dominates the Barker function in terms of variance of ergodic averages. When generating birth-death-swap rates via Poisson thinning as outlined in Section 3.4 we expect these results to still hold optimal.

An interesting prospect is raised, however, when considering whether this birth-death-swap process could be generated with more efficient Poisson thinning bounds. In this case the Metropolis balancing function may not be optimal and other balancing functions may be worth investigating.

\section{Swap moves efficiency experiment details}

To conduct the swap experiment in Section 3.4.2 data were generated as the following
\begin{gather*}
    Y_1 \sim \text{log-Normal}(0,0.5), \\
    Y_2 \sim \text{Exponential}(1), \\
    Y = \min\{Y_1, Y_2\}, \\
    Y_{\text{Censored}} \sim \text{Exponential}(0.5), \\
    Y_{\text{Observed}} = \min\{Y, Y_{\text{Censored}}\}.
\end{gather*}
A single binary covariate was also generated from Bernoulli random variables with $p = 0.5$.

Samplers were run for 10,000 time units generating approximately 10,000 samples with birth-death swap moves occuring with $\Lambda^S + \Lambda^{BD} = 10$.

\section{Additional details for Section 4}

This Section contains descriptions of the sampler settings used for the Lung transplant data, COST data and Kidney transplant data as well as trace plots. Permission to use the lung transplant data was kindly provided by the authors of the original paper. The subset of data used for the COST analysis is available via the R \texttt{pec} package, and the kidney transplant data is available via Dryad as described in the main paper.

\subsection{Lung transplant data}

\begin{figure}
    \centering
    \includegraphics{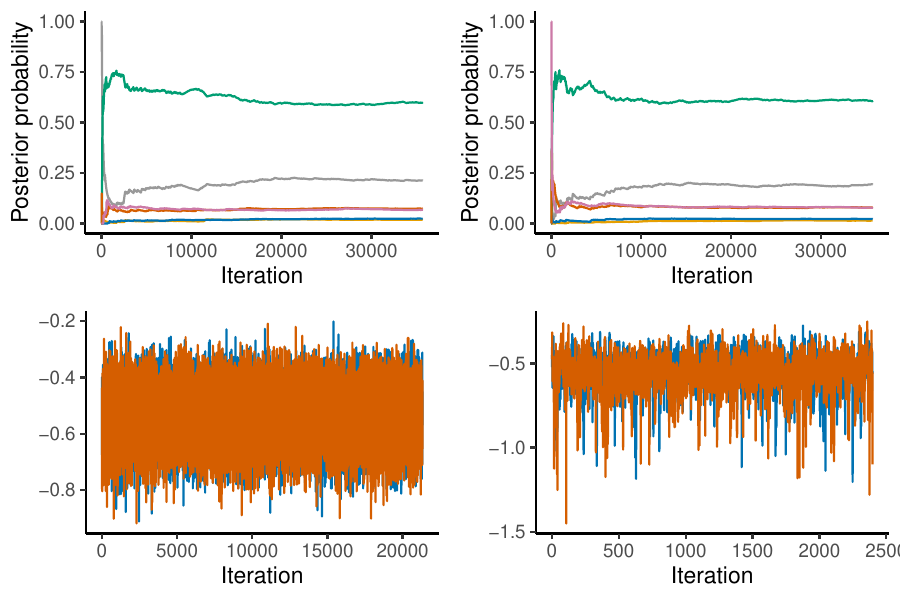}
    \caption{Trace plots for the extended polyhazard model fit to the Lung transplant data for: (First row) Posterior sub-model probabilities from the first and second chain; (Second row) The $\alpha_1$ from the bi- and tri-log-logistic models.}
    \label{fig:LungPlot}
\end{figure}

The extended polyhazard model was fit using the prior specification outlined in Section 2.3. except for the hyperprior specification were fixed hyperparameters were used. The sampler was run for 10,000 time units, with samples taken with rate 4 and $\Lambda^{BD} + \Lambda^S = 10$. Trace plots for submodel posterior probabilities and a subset of parameters are shown in Figure \ref{fig:LungPlot}.

\subsection{COST data}

\begin{figure}
    \centering
    \includegraphics{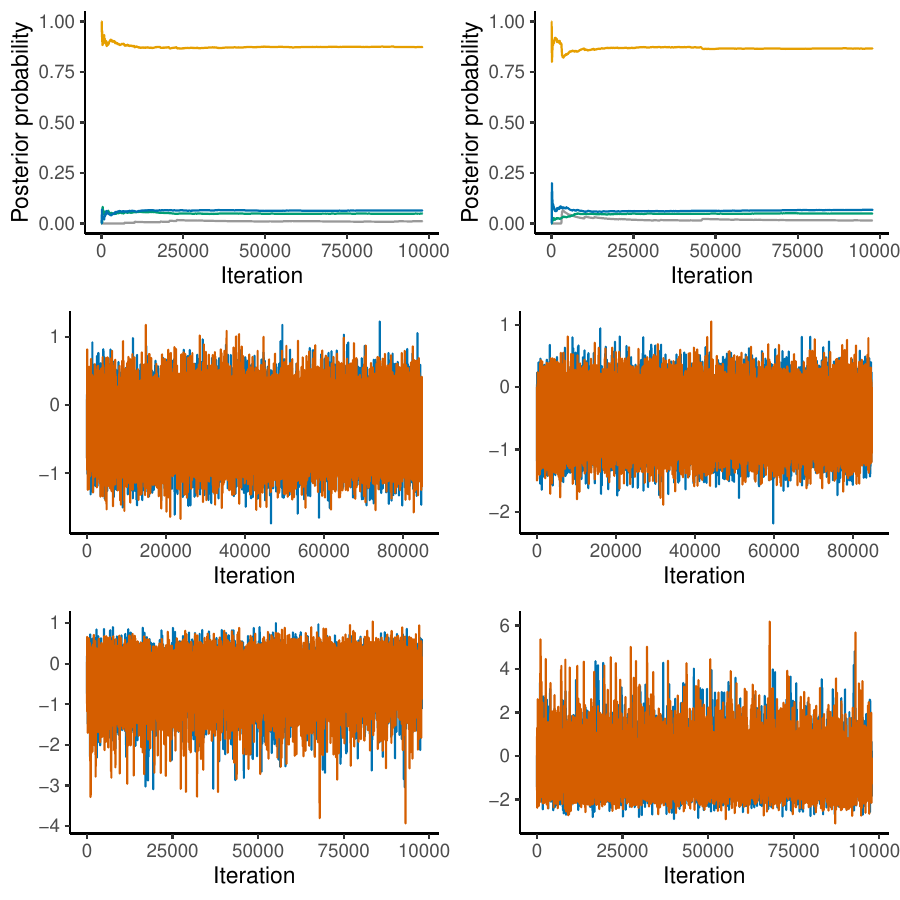}
    \caption{Trace plots for the extended polyhazard model fit to the COST data for: (First row) Posterior sub-model probabilities from the first and second chain; (Second row) Coefficient effects from the bi-log-logistic model; (Third row) $z_1$ and $z_2$.}
    \label{fig:COSTPlot}
\end{figure}

The extended polyhazard model was fit using the prior specification outlined in Section 2.3. The sampler was run for 50,000 time units, with samples taken with rate 5 and $\Lambda^{BD} = \Lambda^S = \Lambda^H = 3.33$. Convergence plots for submodel posterior probabilities along with trace plots for a subset of parameters are shown in Figure \ref{fig:COSTPlot} using three chains.

\subsection{Kidney transplant data}

\begin{figure}
    \centering
    \includegraphics{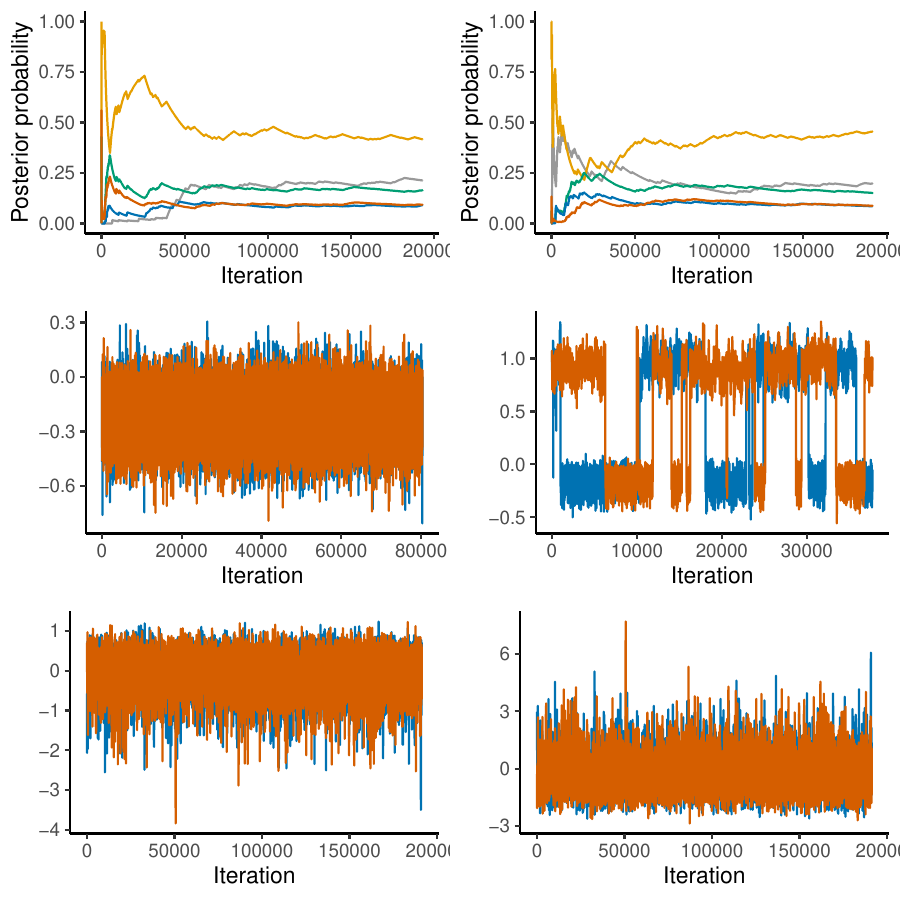}
    \caption{Trace plots for the extended polyhazard model fit to the COST data for: (First row) Posterior sub-model probabilities from the first and second chain; (Second row) A coefficient effect from the bi-Weibull (left) and the Weibull shape parameter from the Weibull-log-logistic (right) models. Note the multi-modality in the model with hazards from different distributions; (Third row) $z_1$ and $z_2$.}
    \label{fig:TKTPlot}
\end{figure}

The extended polyhazard model was fit using the prior specification outlined in Section 2.3. The sampler was run for 10,000 time units, with samples taken with rate 10 and $\Lambda^{BD} = \Lambda^S = \Lambda^H = 6.67$. Convergence plots for submodel posterior probabilities along with trace plots for a subset of parameters are shown in Figure \ref{fig:TKTPlot} using three chains.

%%  imsart-nameyear.bst  will be used to                   %%
%%  MR numbers will be added by VTeX.                      %%
%%                                                         %%
%%  Use \cite{...} to cite references in text.             %%
%%                                                         %%
%%%%%%%%%%%%%%%%%%%%%%%%%%%%%%%%%%%%%%%%%%%%%%%%%%%%%%%%%%%%%

%% if your bibliography is in bibtex format, uncomment commands:
%\bibliographystyle{imsart-nameyear} % Style BST file
%\bibliography{bibliography}       % Bibliography file (usually '*.bib')

%% or include bibliography directly: